# From Signals to Trajectories: A Primer on Low-Dimensional Dynamics in Human EEG and MEG

Vanessa Hadid [1,4], Hamza Abdelhedi[2,5,7], Annalisa Pascarella[3], Sara A. Solla[8,9,10], Paul Cisek[6,11], Karim Jerbi[2,4,5,6]

[1] McGill University Health Centre, McGill University, Montréal, QC, Canada
[2] Cognitive and Computational Neuroscience Laboratory, Université de Montréal, QC, Canada
[3] Institute of Applied Mathematics M. Picone, National Council of Research, Rome, Italy
[4] Psychology Department, Université de Montréal, QC, Canada
[5] MILA (Quebec Artificial Intelligence Institute), Montreal, QC, Canada
[6] Centre UNIQUE (Quebec Neuro-AI Research Center), Montreal, QC, Canada
[7] Pharmacology and Physiology Department, Université de Montréal, QC, Canada
[8] Department of Neuroscience, Northwestern University, USA
[9] Department of Physics and Astronomy, Northwestern University, USA
[10] NSF-Simons National Institute for Theory and Mathematics in Biology, USA
[11] Department of Neuroscience, Université de Montréal, QC, Canada

**Abstract**

Neural activity unfolds not as a set of independent signals, but as a coordinated dynamical process that can be described as a point moving through a high-dimensional state space. This perspective has contributed substantially to recent developments in systems neuroscience, especially through studies of directly recorded neuronal population activity, but remains comparatively underused in non-invasive human recordings such as EEG and MEG. In this primer, we present a neural trajectory analysis based on Principal Component Analysis (PCA) as an accessible route into extending the concepts of neural trajectories in state space to human electrophysiology. We explain how EEG/MEG activity can be organized into neural state representations, projected onto axes organized by decreasing covariance, truncated into low-dimensional representations, and visualized as trajectories that capture how distributed activity evolves over time. We show how trajectory geometry can be used to compare conditions and relate neural dynamics to behavior or clinical outcomes. An example of EEG motor execution and imagery, accompanied by a tutorial notebook, illustrates the workflow from conventional EEG summaries to trajectory-based analysis. We also clarify the limits of PCA, including its linear, variance-driven nature, and discuss EEG/MEG-specific challenges such as spatial mixing, preprocessing sensitivity, validation, and overinterpretation. Finally, we situate PCA within a broader family of state-space methods. By making neural trajectories practical and interpretable, this primer offers a guide for bringing the conceptual framework of population-level dynamics into mainstream human cognitive neuroscience.

## 1 Introduction

Understanding how cognitive processes unfold in time requires analytical and conceptual frameworks that can capture the coordinated activity of large neural populations and follow it over time. In recent years, work in motor and decision neuroscience has increasingly shifted from focusing on individual neurons to emphasizing population-level dynamics. This change has been driven largely by studies that apply dimensionality-reduction techniques to neuronal spiking activity and direct population recordings; these applications have shown that complex computations can unfold along smooth, low-dimensional trajectories embedded within high-dimensional state spaces (Churchland et al., 2012; Shenoy et al., 2013; Cunningham & Yu, 2014; Gallego et al., 2017; Vyas et al., 2020; Jazayeri & Ostojic, 2021; Langdon et al., 2023). These insights have transformed how researchers conceptualize neural computation: not as isolated responses or transient components, but as structured trajectories evolving through low-dimensional state spaces.

Despite the impact of these ideas, their application to human electrophysiology, particularly electroencephalography (EEG) and magnetoencephalography (MEG), remains comparatively limited. This gap is surprising. EEG and MEG provide multivariate, time-resolved measurements that are well suited to geometric state-space representations of neural activity, with millisecond temporal resolution and distributed measurements of brain activity (Baillet, 2017; Cohen, 2014; Gramfort et al., 2013, 2014; Arcara et al., 2023). At each time point within a trial, EEG and MEG simultaneously sample a pattern of activity across sensors, defining a neural state in the original high-dimensional sensor space. The evolution of these states traces a trajectory that can be examined within individual trials or summarized across trials and participants to reveal reproducible condition- and group-level structure. Yet, mainstream EEG/MEG analyses often rely on channel-by-channel averages, scalp maps, oscillatory power measures, or event-related potentials, leaving the underlying population dynamics largely unexamined.

Principal Component Analysis (PCA) is a widely used dimensionality-reduction method for identifying dominant patterns of covariation in multivariate data (Greenacre et al., 2022) while capturing the first and second moments of the data statistics. This Primer focuses on its application to EEG and MEG recordings, providing a practical and conceptual framework for examining how these patterns are expressed over time and how their temporal organization can be represented as trajectories in a low-dimensional state space. PCA offers an accessible entry point to state-space thinking because it projects high-dimensional EEG or MEG measurements onto a smaller set of orthogonal axes, allowing their temporal evolution to be represented within a low-dimensional PCA space.

Although this idea can be applied to any time-series, its conceptual impact has been especially clear in systems neuroscience, where PCA and related low-dimensional approaches have transformed the study of population activity. Studies of population spiking activity in non-human primates during motor control and decision making have shown that complex patterns of neural firing can organize into structured patterns easily revealed in their low-dimensional dynamics. These dynamics reveal population-level motifs that are difficult to detect from single neurons alone or from high-dimensional population state spaces associated with the simultaneous recording of many individual neurons. By capturing preparation, movement, choice formation, context-dependent computation, stable motor manifolds, and decision-related trajectories within a common geometric framework, this work helped shift the field from a language of isolated single neuron responses and tuning curves toward one of evolving population states and trajectories through neural state space (Churchland et al., 2012; Shenoy et al., 2013; Mante et al., 2013; Kaufman et al., 2014; Elsayed et al., 2016; Kobak et al., 2016; Gallego et al., 2017, 2018, 2020; Russo et al., 2020; Vyas et al., 2020; Thura et al., 2022; Langdon et al., 2023; Oby et al., 2025).

The success of these approaches has motivated growing interest in whether similar low-dimensional representations can be used to understand large-scale neural dynamics in human electrophysiology. PCA is already well established in EEG and MEG research, where it has been used for dimensionality reduction, preprocessing before independent component analysis, microstate analysis, feature extraction for machine-learning classification and decoding, source-level summarization before connectivity estimation, and artifact-subspace removal through signal-space projection methods (Uusitalo & Ilmoniemi, 1997; Hyvärinen & Oja, 2000; Kayser & Tenke, 2003; de Cheveigné & Simon, 2007; Artoni et al., 2018; von Wegner et al., 2018; Saeidi et al., 2021; Pellegrini et al., 2023; Brkić et al., 2023; Zhang et al., 2024). Reduced representations obtained through dimensionality reduction are often referred to as latent representations in a latent space. In this Primer, we instead use the terms PCA space or low-dimensional state space to avoid implying that PCA dimensions necessarily correspond to unobserved biological variables.

In most of the applications cited above, PCA serves primarily as a tool for compression, preprocessing, denoising, or feature extraction. In contrast, here we focus on a complementary use of PCA: not simply to reduce dimensionality, but to characterize how multivariate neural activity evolves over time within a low-dimensional state space. This reduced state space provides a compact representation that summarizes the measured multivariate activity, with each principal component defining one axis of the representation. In this framework, the brain activity at each moment in time is represented as a point in state space, and the succession of these points forms a trajectory whose geometry can be quantified and interpreted. Such trajectories provide a way to monitor and analyze neural dynamics directly, offering insights that may not be apparent from individual channels, isolated components, or predefined temporal windows alone. These low-dimensional trajectories provide a compact and intuitive window into the organization of cognitive processes, revealing, for example, how different task conditions separate, converge, or rotate over time within a shared state space. Although only a few human electrophysiology studies have applied these approaches so far, recent work illustrates their potential and added value. In semantic decision-making, low-dimensional EEG trajectories have revealed choice-dependent paths whose location in state space relates to reaction time, suggesting that the evolving neural state carries behaviorally relevant information about the decision process (Do et al., 2024). In studies of attention, low-dimensional state spaces, shared among participants performing the same task, have been used to characterize large-scale neural dynamics associated with cognitive engagement and attentional fluctuations (Song et al., 2023). In MEG, low-dimensional oscillatory trajectories have helped track evolving sensory evidence, choice commitment, and motor execution during dynamic decision-making (Thiery et al., 2022), in a manner similar to what has been shown through neural recordings in monkeys (Thura et al. 2022).

Our goal is to show how this framework can be used by cognitive neuroscientists who may be unfamiliar with state-space approaches yet routinely work with rich multichannel datasets. We review the conceptual foundations of low-dimensional neural state spaces, explain how PCA components, scores, and trajectories can be interpreted in EEG/MEG, and outline practical choices for applying this approach. We also discuss methodological challenges specific to macro-scale electrophysiology and present a roadmap for integrating PCA-based state-space analyses into mainstream EEG/MEG practice.

By bridging classical electrophysiological methods with the population-dynamics toolkit of systems neuroscience, PCA-based neural trajectory analysis has the potential to enrich cognitive neuroscience with new ways of visualizing and quantifying brain computations, bringing human research closer to neural population computational principles observed across species and recording scales.

## 2 Setting the stage for state-space thinking: neural manifolds and trajectories in EEG and MEG

Given that EEG and MEG recordings are, by nature, high-dimensional, brain activity at any given time can be viewed as a coordinated pattern across many sensors, sources, or frequency-specific features. The key intuition behind neural manifold approaches is that brain activity does not necessarily explore this entire high-dimensional space freely. Instead, meaningful neural dynamics may be constrained to a smaller number of coordinated dimensions, forming a lower-dimensional structure embedded within the larger-dimensional empirical space of recorded signals. This structure is what we refer to as a neural manifold. A useful analogy is a dancer moving across a stage: although the body has many joints, and each joint angle could in principle define a separate dimension, a dance does not explore all possible body configurations randomly. Rather, movement draws on a limited vocabulary of coordinated patterns that can be combined and expressed in various manners over time. Neural activity can be understood in a similar way: although EEG and MEG measure many simultaneous signals, task-relevant activity may be described by a smaller set of dominant patterns.

Dimensionality reduction methods aim to reveal this structure. Among them, PCA provides a simple and intuitive entry point by identifying the main directions along which multichannel activity covaries. PCA does not uncover independent neural sources or causal mechanisms; rather, it offers a linear view of the dominant statistical structure in the data. Once EEG or MEG activity is projected onto a few principal components, the instantaneous collective activity can be represented by coordinates in a simplified neural state space. A neural state is the brain's activity pattern at that moment, and the state space is the geometric space in which such states can be represented. As time unfolds, successive states can be connected into a neural trajectory: the path followed by brain activity during perception, decision-making, memory, action, or rest. This is the essence of state-space thinking: representing brain activity as a succession of multivariate states and asking how these states evolve over time. This perspective shifts the question from whether activity increases or decreases at a particular channel to how the whole multivariate pattern evolves. Used carefully, PCA-based state-space analysis offers an accessible bridge between traditional EEG/MEG analysis and the population-dynamics framework that has significantly reshaped systems neuroscience.

## 3 Mathematical foundations of PCA-based neural state spaces

Before applying PCA, each feature is mean-centered across the observations used in the PCA, so that its values represent deviations from its mean. PCA takes the original N-dimensional recorded signals - such as sensors, sources, or frequency features – and constructs a state space in which each one of these signals is a coordinate axis. PCA then finds the direction of highest variance within the N-dimensional state space and assigns this direction as Principal Component 1 ($PC_1$). It then projects the data onto the (N−1)-dimensional hyperplane perpendicular to $PC_1$, and finds the next direction of highest remaining variance, assigning that as Principal Component 2 ($PC_2$). This process is repeated until all N dimensions have been considered or until a desired level of variance has been accounted for (Jolliffe & Cadima, 2016). The result is a set of axes ordered by variance explained; the first few often capture the main features of the data in a way that can be usefully interpreted. In fact, even with very high dimensional data sets one often finds that 80-90% of the total variance can be accounted for with something like 5-7 PCs, which can then be the focus of subsequent analyses. Each signal can be mapped to a given PC through a set of weights, so that the time course of the N-dimensional original system can be followed in the subspace defined by the selected dominant PCs. The sign of each principal component is arbitrary: multiplying all its weights, and consequently its scores as defined below, by −1 yields an equivalent component and does not change the underlying geometry.

For a multichannel recording, the score of Principal Component $k$ at time $t$ can be written as a weighted sum of the original signals:

$$PC_k(t) = w_{1k}x_1(t) + w_{2k}x_2(t) + \ldots + w_{Nk}x_N(t)$$

Here, $x_i(t)$ is the signal from channel, voxel, or feature $i$, and $w_{ik}$ is the weight of that feature on $PC_k$. These weights define the spatial or feature pattern associated with the component, while the score describes how active the *k* component is, i.e., how strongly this pattern is expressed at a given time *t*. In EEG and MEG, these weights can be visualized as scalp topographies or source-level maps, linking the low-dimensional trajectory back to the original electrophysiological signals (Kayser & Tenke, 2003; Gramfort et al., 2013, 2014). The resulting PCA scores therefore provide a low-dimensional representation of the measured multivariate activity.

As mentioned, neural trajectories are obtained by ordering PCA scores over time. If only one component is retained, $PC_1(t)$ describes a one-dimensional trajectory along the first principal axis. With two or three retained components, each time point is represented as a coordinate in a two- or three-dimensional space, such as ($PC_1(t)$, $PC_2(t)$) or ($PC_1(t)$, $PC_2(t)$, $PC_3(t)$). Connecting these successive points in temporal order describes how the multichannel EEG or MEG pattern evolves in the reduced dimensionality state space. Separate trajectories can then be obtained for trials, conditions, subjects, or group averages, depending on how the data were organized before PCA and projected afterward.

PCA implies a direct linear relationship between the original data and the reduced dimensionality state space, making it an accessible bridge between classical electrophysiological analysis and state-space approaches. However, PCA also has important limits. It is linear, variance driven, and sensitive to scaling and preprocessing. The components that explain the most variance are not necessarily the components most relevant to a task effect, behavioral variable, or clinical difference. PCA also does not separate independent neural sources or solve the inverse problem. In EEG and MEG, where sensors measure spatially mixed activity, PCA organizes the measured signal according to covariance structure rather than anatomical or causal origin (Makeig et al., 1996; Jung et al., 2001; Baillet, 2017).

## 4 Constructing low-dimensional neural state spaces

The construction of low-dimensional neural state spaces involves several decisions that shape the resulting geometry. The first is the definition of the empirical neural state. This is typically a high-dimensional space, based on the quantities that are being measured. In sensor space, each state may correspond to the pattern of activity across EEG or MEG sensors at a given point in time. In source space, the state may correspond to activity across cortical regions defined by an anatomical or functional parcellation. Time–frequency analyses can define states using the spatial distribution of band-limited power or other spectral features across sensors, sources, or regions. Connectivity-based trajectories may describe changing relationships among regions. These alternatives differ both in the neural representation they emphasize (e.g., voltage, spectral power, or connectivity) and in the spatial scale at which activity is described (e.g., sensors, sources, or regions). The choice of representation should therefore be guided by the scientific question being addressed (Cohen, 2014; Morales & Bowers, 2022; Pellegrini et al., 2023).

The second decision concerns preprocessing. Because PCA emphasizes sources of large variance, residual artifacts, slow drifts, high-amplitude sensors, line noise, or unbalanced trial numbers can influence the PCA space. Filtering, artifact correction, baseline correction, re-referencing, epoching, source reconstruction, and normalization should therefore be reported clearly and tested for robustness (Keil et al., 2014). Scaling may be useful when the goal is to prevent high-amplitude channels or subjects from dominating the analysis, but it should not be applied automatically if amplitude differences are conceptually meaningful to the question being addressed.

The third decision concerns the construction of the data matrix. EEG and MEG data are typically multidimensional and can be organized as channels × time × trials × conditions × subjects. For trajectory analysis, the temporal dimension must be preserved so that trajectories can be reconstructed from successive time points in the reduced state space. Depending on the analysis goal, these time-indexed observations may correspond to single-trial samples, condition-averaged responses, subject-level averages, or pooled observations across subjects and conditions. Columns define the N features used to describe each neural state, such as sensors, sources, frequencies, or connectivity values. This organization determines what PCA captures and how the resulting components and trajectories should be interpreted.

A related decision is how to apply PCA to the data. PCA can be fit within individual subjects, within specific conditions, at the group level, within a specific task epoch, or within an ROI or network. When PCA is fit separately for each subject or each condition, the resulting PC axes are estimated independently and are therefore not directly comparable across participants or conditions. This approach can be useful for characterizing within-subject or within-condition dynamics, but it limits direct comparison of trajectory geometry across datasets. In contrast, fitting PCA at the group level allows all subjects and conditions to be projected into a common coordinate system, although the representation of individual-specific structure may be obscured. For direct comparisons between conditions, subjects, or groups, a shared PC space is required so that all observations are expressed within the same axes. Table 1 summarizes common PCA strategies and the trajectory-analysis levels that naturally follow from them.

| PCA strategy | Use when the goal is to... | Appropriate trajectory analysis level | Main caution |
|---|---|---|---|
| Within-subject PCA* | Characterize individual neural dynamics or within-subject effects. | Single-trial trajectories; condition-averaged trajectories; subject-level trajectory metrics. | Useful for individual-level analyses, but not for direct comparison of coordinates across participants. Sign, order, and orientation of PCs may vary across individuals. |
| Across-condition PCA within subject* | Compare conditions for a given participant in a shared subject-specific space. | Condition-averaged trajectories; within-subject separation metrics; trial-wise condition effects. | Avoid fitting separate PCAs per condition if the goal is direct condition comparison, because each condition would then have its own coordinate system. |
| Shared PCA-space projection | Define one common PCA space, often from pooled participants, trials, or conditions, and project subject-, trial-, or condition-level data into that same coordinate system. | Single-trial trajectories; subject-level trajectories; condition averages; group-level condition comparisons; metric-level group statistics. | The shared PCA space must be fit on data appropriate for the planned comparison. Avoid fitting the PCA on data that has already averaged away the variability needed for the analysis. For decoding or cross-validation, PCA should be fit only on the training data. |
| Task- or epoch-specific PCA | Focus on a specific task phase, such as encoding, maintenance, decision, or response preparation. | Epoch-specific trajectories; time-windowed metrics; condition separation during selected periods. | The resulting space may not generalize to other task periods. If the goal is to compare phases, a common space across phases is required. |
| ROI- or network-specific PCA | Examine localized or network-level dynamics rather than whole-brain or whole-sensor patterns. | ROI trajectories; network-level metrics; subject- or condition-level comparisons. | May miss distributed interactions outside the selected region or network. Results depend on how ROIs or networks are defined. |

**Table 1 | Choosing the PCA space and trajectory-analysis level.** PCA defines the coordinate system in which trajectories are represented, whereas trajectory analysis determines how movement through that space is visualized, summarized, and tested. Different strategies are appropriate depending on whether the goal is to study individual dynamics, condition effects, group-level structure, trial-to-trial variability, or trajectory metrics. In many EEG/MEG applications, a useful approach is to define a shared PCA space and then project trials, conditions, or subjects into that space so that their trajectories can be compared directly. *When PCA is fit separately for each subject, the resulting axes are subject-specific and are not directly comparable across participants. This applies both to within-subject PCA and to across-condition PCA performed within each subject.

Once PCA is complete, the data can be projected into the selected components. For visualization, two or three components are often used. For inference, however, the number of retained components should be justified by variance accounted for, component stability, interpretability, and the robustness of downstream trajectory metrics. Lower-variance components should not be dismissed automatically, because task-relevant or behaviorally relevant information may appear outside the first few components (Cunningham & Yu, 2014; Yan et al., 2020; Heller & David, 2022). Figure 1 summarizes this workflow as a ten-step guide for deriving and interpreting PCA-based neural trajectories from EEG/MEG data.

Importantly, once PCA has produced the set of weights that map each signal $x_i$ to each component $PC_k$, that same set of weights can be used to project any data from those signals to the same state space, even data that was not used to perform the original PCA analysis. For example, one could perform PCA on signals from a subject performing a simple behavioral task, obtain a set of components that define a transformed state space, and then calculate what the trajectories look like in the transformed state space when the subject performs a completely different task.

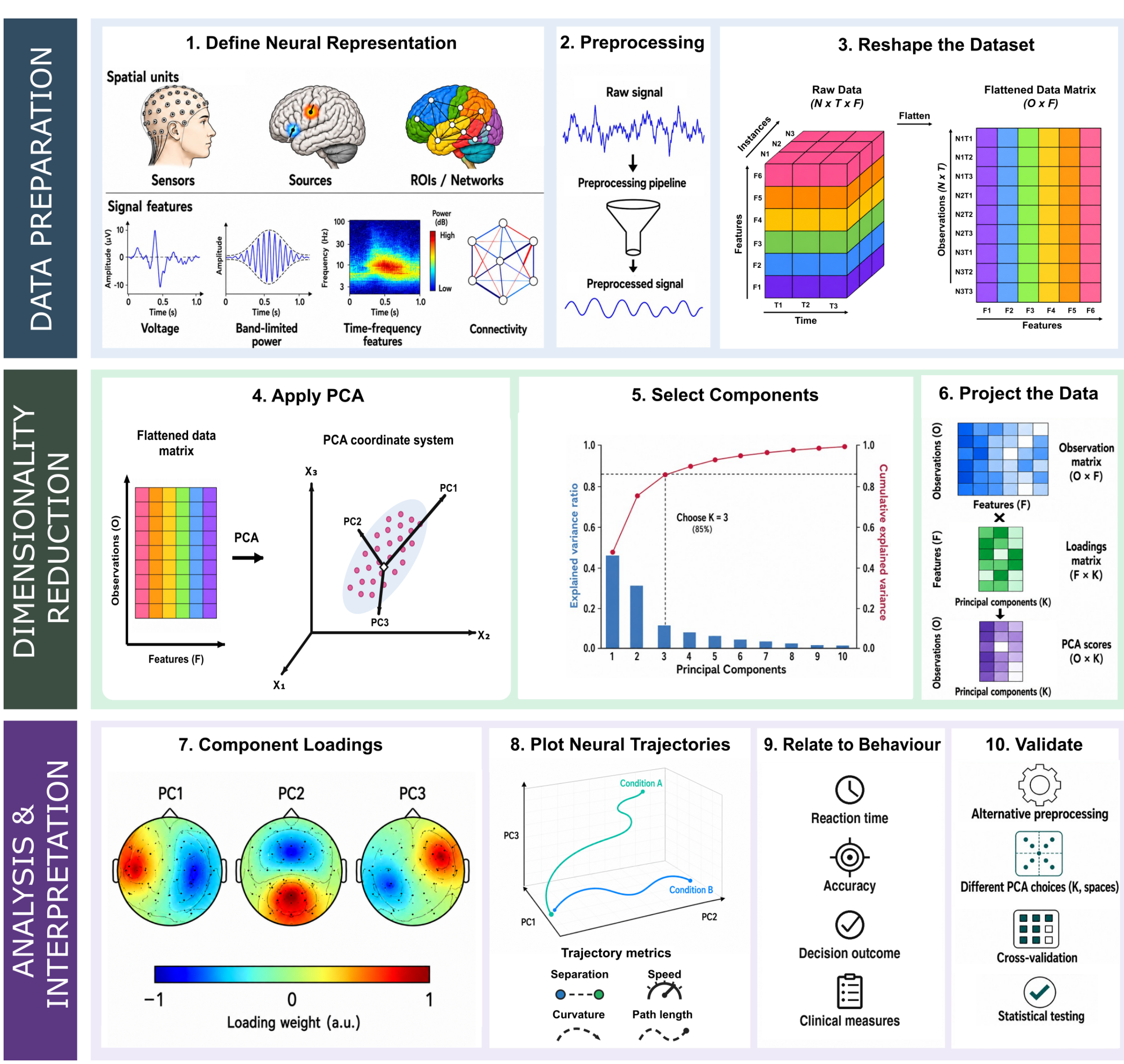

**Figure 1 | PCA-based neural trajectory analysis from data representation to dynamical interpretation.** The workflow is organized into three stages: data preparation, dimensionality reduction, and analysis and interpretation. **1. Define the neural representation.** Select the spatial representation of the data, such as sensors, sources, or ROIs/networks, and the signal features to analyze, such as time-domain voltage or ERPs, band-limited power, time–frequency features, or connectivity. The value of the selected features at each observation form a high-dimensional neural state vector; the space in which these vectors are defined is the original state space. **2. Preprocess the data.** Apply an appropriate preprocessing pipeline to reduce noise and

artifacts while preserving the neural signals of interest. **3. Reshape the dataset.** Arrange the data into an observation-by-feature matrix, $\boldsymbol{X} \in \mathbb{R}^{O \times F}$. Observations may represent time points across trials, participants, or conditions, whereas features may represent sensors, sources, voxels, ROIs, connections, or other signal dimensions. When time points are treated as observations, PCA itself is agnostic to their temporal order; temporal structure is reintroduced when the projected scores are ordered in time to form trajectories. **4. Apply PCA.** PCA identifies orthogonal loading vectors in the original $F$-dimensional feature space. These loading vectors define the principal-component axes, which are ordered according to the amount of variance they explain, with $PC_1$ explaining the greatest variance, followed by $PC_2$, $PC_3$, and subsequent components. **5. Select components.** Choose the number of retained components $K$, based on explained variance, interpretability, and the goals of the analysis. Measures of effective dimensionality, such as the participation ratio, provide a complementary description of how variance is distributed across components and whether the data are concentrated within a small number of dimensions. **6. Project the data.** Multiply the observation matrix $\boldsymbol{X}$ by the loading matrix $\boldsymbol{W}_K \in \mathbb{R}^{F \times K}$ to obtain the score matrix, $\boldsymbol{Z} = \boldsymbol{X}\boldsymbol{W}_K$, which represents each observation in the reduced $K$-dimensional PC space. In the example shown here, $K = 3$. **7. Examine component loadings.** Inspect the loading vectors to determine how the original features contribute to each retained component. When the features correspond to sensors, these weights can be visualized as scalp topographies. Their magnitude and sign describe relative contributions within each component, although the overall sign of a PC is arbitrary. **8. Plot neural trajectories and quantify their geometry.** Connect time-ordered PC scores to visualize how neural states evolve through the reduced space and compare trajectories across experimental conditions. Quantify trajectory geometry using measures such as between-condition separation, speed, curvature, and path length. **9. Relate neural measures to behaviour and outcomes.** Test whether component-loading patterns or trajectory-derived metrics are associated with behavioural or clinical outcomes, including reaction time, accuracy, decision outcomes, and clinical measures. **10. Validate and refine.** Evaluate the robustness of the findings using alternative preprocessing choices, different PCA spaces or values of $K$, alternative trajectory metrics, cross-validation, and appropriate statistical testing.

## 5 Quantifying and interpreting neural trajectories

Several metrics can be used to characterize trajectories as described in Table 2. Speed, or state-space velocity, captures how quickly the projected neural state changes over time. Curvature describes how sharply the trajectory bends and may help identify transitions between processing stages. Path length summarizes the total movement through projected state space during a trial, condition, or task epoch. Separation measures how far apart condition-specific trajectories are, while convergence and divergence describe whether trajectories move toward common or distinct regions. Subspace overlap, alignment, and dimensionality estimates can test whether conditions, participants, tasks, or groups rely on similar low-dimensional structures (Chung & Abbott, 2021; Recanatesi et al., 2022). Beyond these comparisons, the geometry of the subspace occupied during a particular behavioral state may also provide insight into how the contributing neural populations are coordinated during that state.

These measures connect EEG/MEG trajectory analysis with neuro-AI, where related geometric tools are used to compare representational structure and dimensionality in biological and artificial neural networks. Speed may be relevant when studying reaction time or rapid state transitions. Trajectory separation may be useful when the hypothesis being tested concerns category differentiation, decision formation, or condition-specific representations. Curvature may help characterize nonlinear transitions between task phases. Dimensionality may relate to task complexity, cognitive flexibility, or clinical variability. However, no trajectory metric has an intrinsic cognitive meaning. A bend, loop, separation, or attractor-like pattern becomes interpretable only when linked to task timing, behavior, component structure, robustness analyses, or computational models (Sussillo, 2014; Vyas et al., 2020; Mitchell-Heggs et al., 2023).

| Metric | What it measures | Interpretation in neural trajectories | Typical applications |
|---|---|---|---|
| Speed / state-space velocity | Magnitude of distance between successive points | How quickly the neural state evolves over time | Reaction-time correlations; rapid transitions; commitment points |
| Acceleration / jerk | Rate of change of speed or higher-order change | Abrupt shifts or changes in trajectory dynamics | Detecting state transitions or decision boundaries |
| Curvature | Degree to which the trajectory bends | Nonlinear transitions or turning points between processing stages | Distinguishing gradual accumulation from abrupt changes |
| Path length | Total distance traveled in state space | Overall amount of neural change across a trial or condition | Quantifying cognitive effort, representational change, or task demands |
| Trajectory separation | Distance between trajectories for different conditions | Degree to which neural states differentiate categories, decisions, or task rules | Condition comparisons; decoding; representational analyses |
| Divergence / convergence | Whether trajectories move apart or toward common regions | Differentiation versus stabilization of neural states | Category formation; evidence accumulation; attractor-like dynamics |
| Trajectory similarity | Similarity between paths, including timing and shape | Degree to which trajectories follow comparable routes through state space | Comparing subjects, trials, or conditions |
| Subspace overlap | Degree to which trajectories occupy similar low-dimensional subspaces | Shared versus task-specific neural dynamics | Comparing tasks, groups, modalities, or models |
| Dimensionality | Effective number of axes needed to describe the activity | Complexity of the underlying neural dynamics | Task difficulty, learning, clinical comparisons, model comparison |
| Trajectory alignment | Similarity of temporal structure across subjects or sessions | Cross-participant or cross-session consistency of dynamics | Group-level analyses; longitudinal analyses; normalization |

**Table 2 | Metrics for quantifying neural trajectories.** Summary of commonly used geometric and dynamical metrics for analyzing trajectories in low-dimensional neural state spaces. These measures capture complementary aspects of trajectory evolution, including speed, curvature, path length, condition separation, dimensionality, subspace alignment, and stability. Together, they provide quantitative tools for comparing neural dynamics across time, conditions, individuals, and behavioral outcomes.

# 6 Examples

## Example 1: Motor execution and imagery

We used a motor execution and imagery dataset to illustrate how PCA can bridge conventional electrophysiological summaries and state-space analysis (Figure 2). The data were drawn from the publicly available PhysioNet EEG Motor Movement/Imagery Dataset (Schalk et al., 2004; Goldberger et al., 2000), which comprises 64-channel EEG recordings from 106 subjects performing multiple motor execution and imagery tasks. Here we focus on four conditions: left- and right-hand motor execution and left- and right-hand motor imagery. Each trial lasted approximately 4 s, with cues presented visually. Full preprocessing details, including bandpass filtering (0.5–40 Hz), ICA-based blink and heartbeat removal, and epoching, are described in the companion tutorial notebook. The data are organized as a 4D tensor of dimensions T x N x S x N_ch where T = 193 is the number of time samples per epoch (−0.2 to 1.0 s at 160 Hz), N = 4 the number of conditions, S = 106 the number of subjects, and N_ch = 64 the number of EEG channels (cf. Figure 2).

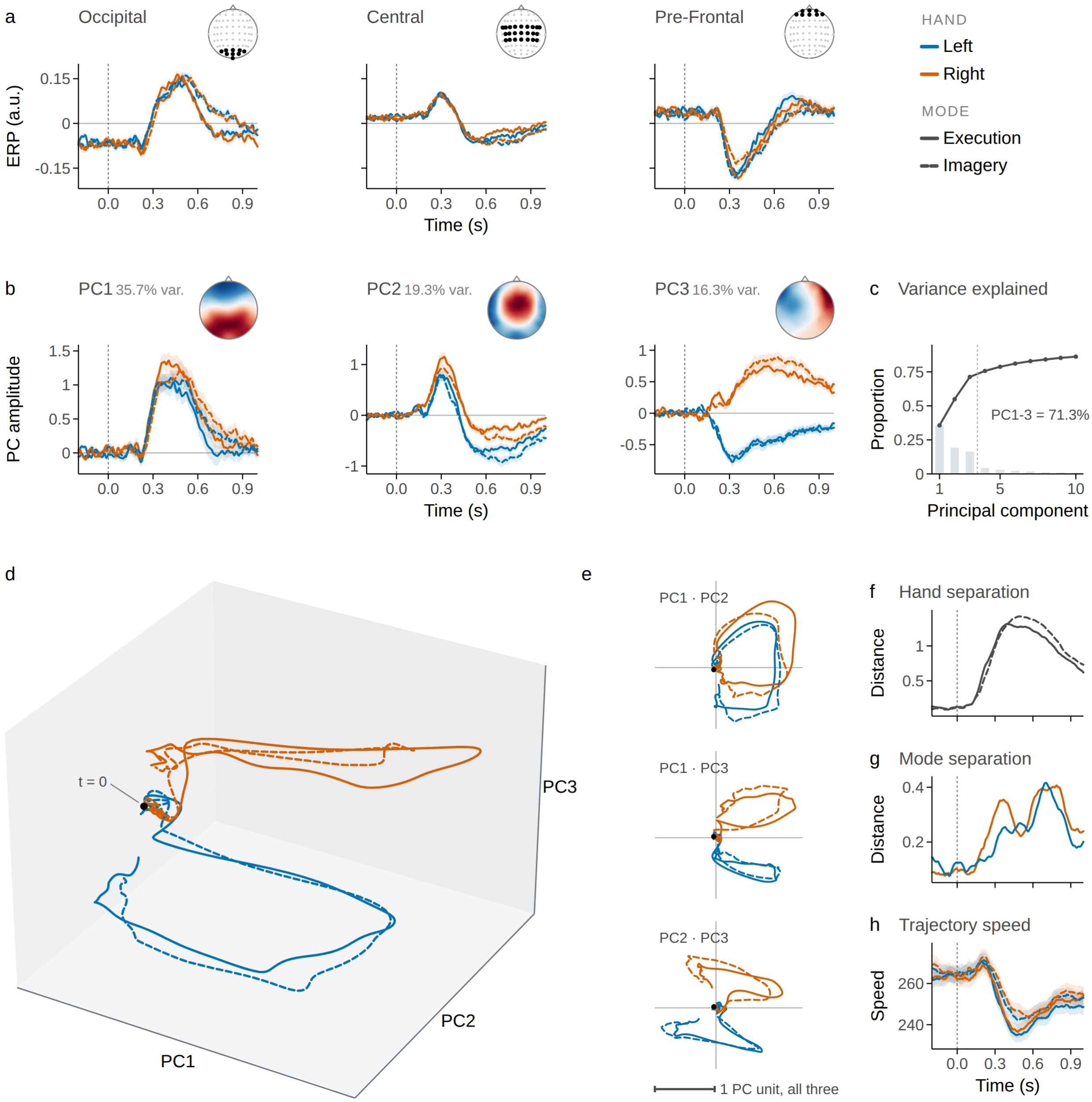

**Figure 2 | Example of PCA-based trajectory analysis in a motor execution and imagery task.** a. Regional ERPs show condition-specific temporal activity for representative prefrontal (8 channels), central (23 channels), and occipital (9 channels) channel groups. Curves show the mean ± SEM across trials pooled across subjects within each condition. b. The first three principal components are shown as time courses (mean ± SEM across subjects) obtained from the condition-averaged ERPs, with inset scalp maps indicating the corresponding component loadings $w_{ik}$ (see Eq. 1). c. The scree plot shows the variance explained by individual components and the cumulative variance captured as additional components are included in the

shared ERP PCA. d. For each subject and condition, trials were first averaged to obtain a 64-channel ERP that was projected into the shared three-dimensional PC state space. The trajectories shown represent the mean across subjects for each condition and trace the temporal evolution of the projected multichannel EEG pattern. The black filled circle marks t = 0. e. Two-dimensional projections show the same averaged trajectories of panel d across pairwise PC axes, allowing visual inspection of separation, overlap, and trajectory geometry. f. Time-resolved Euclidean distances quantify separation between condition centroids in the shared single-trial PC space. g. Additional Euclidean-distance comparisons derived from the same single-trial PCA summarize separation between execution and imagery conditions across hand conditions. h. Trajectory speed quantifies the rate of movement through the shared single-trial PC space over time. Speed is computed separately for each trial; curves show the mean ± SEM across trial-level speeds pooled across subjects within each condition.

Standard EEG analyses typically begin with sensor-level summaries, including regional ERPs, condition-specific time courses, and scalp topographies. Here, regional ERPs were averaged across representative occipital (9 channels), central (23 channels), and prefrontal (8 channels) channel groups (Figure 2a), comprising 40 of the 64 recorded channels, with t=0 corresponding to cue onset. These conventional averaged outputs are useful because they show when activity changes, where the strongest sensor-level effects appear, and whether responses differ across left- and right-hand execution or imagery conditions. However, they do not directly reveal how the full multichannel pattern evolves over time as a coordinated neural state.

For the group-level PCA representation, trials were first averaged within each condition for each subject, yielding one 64-channel ERP time course per subject and condition. These subject- and condition-specific time courses were then concatenated across time points, conditions, and subjects to form a matrix of dimensions (T × N × S) × N_ch, where N_ch = 64 and each row represents the multichannel activity pattern at a particular time point for a given subject and condition. The principal axes were then estimated from this matrix, and the data were projected onto the first K PC axes, yielding a reduced representation of dimensions (T × N × S) × K (cf. Figure 2).

Because the principal axes are estimated from all subjects and conditions simultaneously, the resulting state space provides a common low-dimensional coordinate system for comparing neural dynamics across participants and task conditions. By identifying dominant axes of covariance, PCA transforms the original high-dimensional (N_ch=64) signal into a lower-dimensional (K=3) space. The temporal evolution of the data can be visualized as trajectories in the low-dimensional PC space. The component time courses (Figure 2b)—reveal how strongly each principal component is expressed over time for each condition, while the component loadings $w_{ik}$ (see Eq. 1)—visualized as scalp topographies—link the abstract low-dimensional axes back to the original physical sensor space. The scree plot indicates how much variance is captured by successive components, helping to justify whether a small number of components provides a useful approximation of the original signal—70.3% of the variance is accounted for when using K=3 (Figure 2c).

This link between the PCA space and the sensor space is essential for interpretation. A trajectory is only meaningful if the axes that define it are understood. For example, a component that contributes strongly to condition separation may reflect task-relevant neural activity in the motor cortex, but it may also reflect residual ocular, muscular, or other non-neural artifacts. The component loadings $w_{ik}$ therefore help determine whether the observed trajectory structure is likely to reflect the neural process of interest, a residual artifact, or a mixture of both. In EEG and MEG trajectory analysis, this step should be treated as a mandatory part of the interpretation rather than as a secondary diagnostic.

Once the data are projected into the selected PC space, each condition can be represented as a continuous path through state space. Three- (Figure 2d) and two-dimensional (Figure 2e) trajectory plots—each showing the grand average across subjects, with the black dot marking time t = 0 (cue onset)—make it possible to examine whether motor execution and imagery conditions follow similar or different paths, whether left and right conditions separate, and whether trajectories converge or diverge during specific task periods. Note the crucial role of $PC_3$ in separating left- and right-hand conditions; this brain-wide activity pattern is associated with a left-right asymmetry in neural activity not captured by the original sensor-averaged variables.

We quantified these visual trajectories using trajectory-derived measures. By computing the time-resolved Euclidean distance between specific condition pairs, we quantified the exact timing and magnitude of the separation between conditions (Figure 2f, g). Furthermore, by calculating speed along the trajectory (the rate of movement through state space) at the single-trial level before averaging, we captured how rapidly the multivariate EEG state changes during different task phases (Figure 2h). The statistical significance of these observations can be calculated using a variety of approaches. For example, one can use bootstrapping to estimate the 95% confidence interval around each trajectory by calculating a distribution of 1000 trajectories that each takes a randomly resampled (with replacement) set of the original observations and projects it through the loading matrix $W_K$.

In this way, PCA trajectories extend the analysis from asking whether activity differs at a given sensor to asking how the global neural state unfolds dynamically in different conditions. This example also highlights why visual separation alone is not sufficient evidence for a meaningful neural effect. A trajectory plot can show that two conditions occupy different regions of the state space, but the interpretation of this separation depends on several methodological choices: whether PCA was fit on trial-averaged or single-trial data, how many components were retained, whether the same PCA space was used across all conditions, and whether metrics like speed were computed before or after averaging across conditions. These details determine whether trajectory geometry reflects stable task-related structure, subject- or trial-level variability, preprocessing choices, or residual artifacts. Accordingly, Figure 2 should be read as a workflow example rather than as a standalone demonstration that trajectory separation is intrinsically meaningful.

## Example 2: Face Perception in MEG

Our second example shows that the same workflow applies to a different recording modality, a different definition of the neural state, and to prediction across participants. We used the Wakeman–Henson face-perception MEG dataset (Wakeman & Henson, 2015); it follows 16 participants as they viewed famous, unfamiliar, and scrambled faces. Rather than using the whole sensor array, we restricted the analysis to a predefined 36-sensor right-occipital selection, chosen to emphasize the well-characterized occipitotemporal face response near 170 ms (Liu et al., 2002; Rossion et al., 2003). Because magnetometers and planar gradiometers carry different physical units, the data were noise-whitened with a subject dependent empty-room covariance estimate before PCA.

As in the EEG example, the conventional starting point is a sensor-level summary: the root-mean-square response across the whitened right-occipital sensors begins to separate the three conditions shortly after stimulus onset (Figure 3a). We then built a shared PC state space, as in the EEG example, but performed PCA on the single-trial MEG data rather than on the trial-averaged ERPs. Trials were pooled across participants and conditions without averaging, giving a matrix of dimensions (T × N_tr) × N_ch, where T = 201 is the number of time samples in the fitting window (−0.2 to 0.6 s at 250 Hz), N_tr = 12,642 the number of retained trials across all participants and conditions, and N_ch = 36 the number of sensors. Each row is therefore the multichannel activity pattern at one time point in one trial for a specific condition and participant; condition and participant labels are ignored during PCA fitting. The principal axes were estimated from this matrix, and the full data set was projected onto these axes; the first three components are displayed in Figure 3b. Condition trajectories are formed after projection, by averaging trials within each participant and then across participants, so that each participant contributes equally. As in the EEG example, each point on a trajectory is the projected multivariate MEG state at one moment and the connected path traces how that state evolves through the state space in time. As in the EEG example, the component loadings tie the abstract PC axes back to the sensors; here, the a priori restriction to a right-occipital selection of sensors constrains what the axes can express, so the state space is posterior by construction rather than by inference.

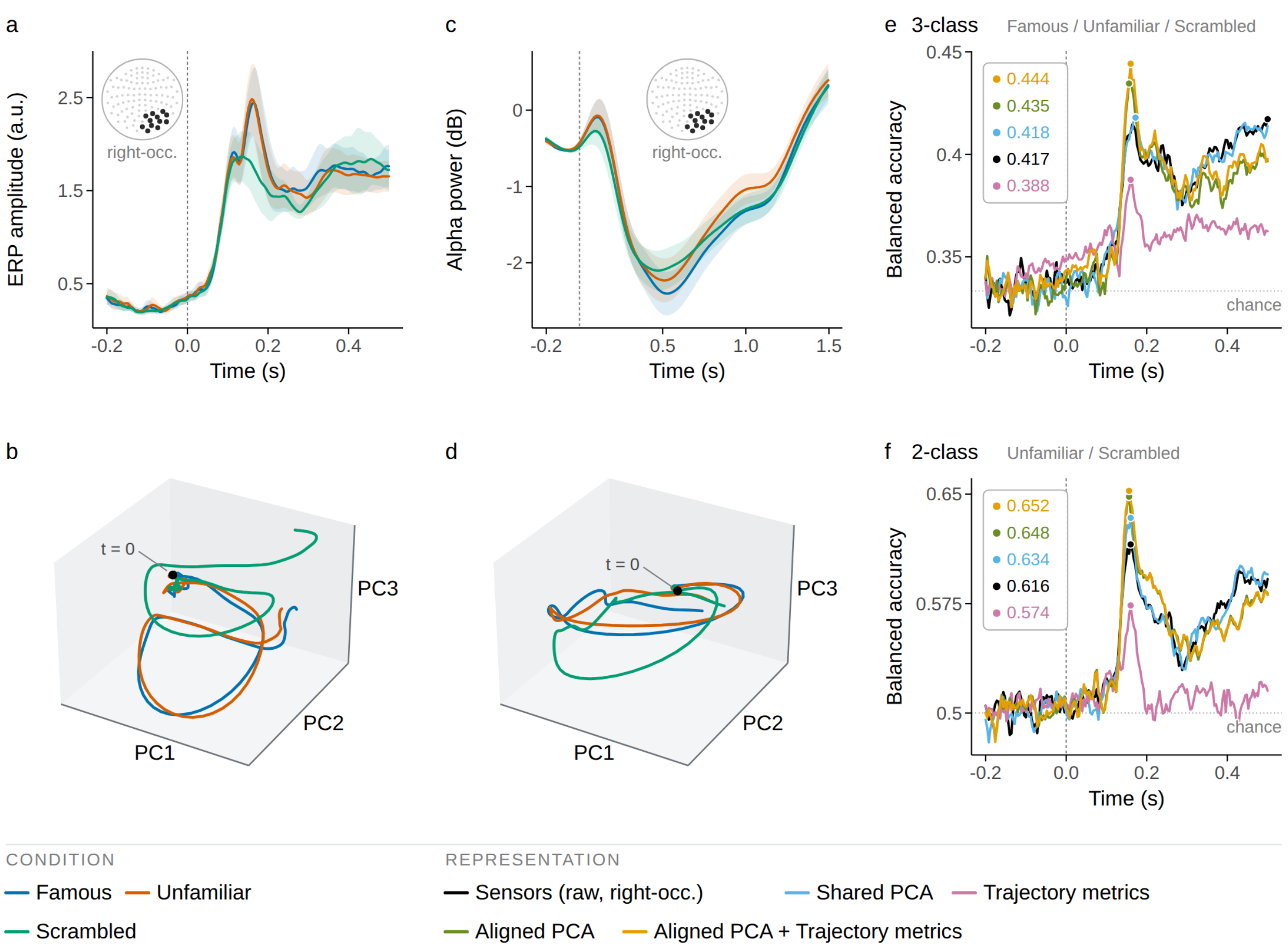

**Figure 3 | PCA trajectories and cross-participant decoding in a face-perception MEG task.** Data are from the Wakeman–Henson face-perception dataset (Wakeman & Henson, 2015); 16 participants viewed famous, unfamiliar, and scrambled faces. All panels use a predefined 36-sensor right-occipital MEG selection. a. Whitened root-mean-square response across the right-occipital sensors (mean ± SEM across participants) for each of the three conditions. The trial-averaged ERP is shown here for visualization; PCA was performed on the corresponding single-trial MEG data. b. Broadband activity in the shared PC state space, shown as the first three of ten retained components, averaged within participants and then across participants; the black marker denotes the common starting point of the three trajectories at t=0. c. Hilbert-transformed alpha-band power (8–12 Hz) in dB relative to the prestimulus baseline (mean ± SEM across participants) for each of the three conditions. d. Alpha-power activity in its own separately fitted PCA state space, plotted with the conventions of panel b. e. Cross-participant decoding of famous, unfamiliar, and scrambled stimuli (chance = 1/3). Curves give time-resolved balanced accuracy under leave-one-participant-out cross-validation for sensor space (black), shared PCA (blue) , trajectory descriptors alone (red), aligned participant-specific PCA (green), and aligned PCA augmented with descriptors (yellow). f. As in panel e, for unfamiliar versus scrambled stimuli (chance = 0.5). In both e and f, values in the inset indicate maximum value for each category within the displayed interval.

To extend the analysis to a different characterization of the neural state, we use a Hilbert transform to compute power in the 8–12 Hz band as the state vector (Figure 3c). These data were projected into their own, separately fitted PC state space (Figure 3d). The resulting trajectories differ from the broadband ones, and that difference is the point: changing the choice of variables that characterize the neural state changes the geometry of the trajectory, although the analysis framework stays the same.

The final step moves from visualization to prediction. We decoded stimulus condition with a sliding-window classifier, using leave-one-participant-out cross-validation across the 16 participants. We compared three representations of each trial: the 36 whitened sensors themselves, the shared PC state space described above, and a set of explicit trajectory descriptors computed from the projected path (speed, jerk, cumulative path length, distance from centroid, turning angle, and the area under the speed curve, among others). Sensor space and shared PCA performed almost identically. Within the early post-stimulus interval shown in Figure 3e–f, balanced accuracy reached 0.417 for sensors and 0.418 for shared PCA in the three-class problem (chance 0.333), and 0.616 and 0.634 for unfamiliar versus scrambled stimuli (chance 0.500). The reduction in dimensionality is essentially free, but it neither buys nor loses prediction accuracy. Trajectory descriptors on their own were clearly worse: 0.388 and 0.574.

We then asked whether aligning the state spaces across participants would help. In aligned PCA, each participant gets their own PCA state space rather than participating in the construction of a shared one. These separately estimated PC spaces are then expressed in a common coordinate system — an approach established for neural population dynamics across sessions and individuals (Gallego et al., 2020; Safaie et al., 2023). Within each cross-validation fold, trials are first averaged across conditions within each training participant to obtain a condition-agnostic mean trajectory. These participant-level trajectories are then averaged across the training participants to define a common template. Each participant's own condition-agnostic mean trajectory is Procrustes-rotated onto that template. This makes participant-specific state spaces comparable. With 22 components, aligned PCA reached 0.435 in the three-class classification and 0.648 for unfamiliar versus scrambled stimuli, outperforming the sensor-space baseline in both cases. Both peaked near 156 ms, close to the expected M170 latency. Finally, we concatenated the trajectory descriptors onto the aligned component scores. The best configurations reached 0.444 in the three-class task (22 components, with path length, displacement, and area under the speed curve) and 0.652 for unfamiliar versus scrambled stimuli (22 components, with jerk, path length, and area under the speed curve). Taken together, the example shows how the PC-trajectory framework extends across signal representations and from visualization to cross-participant prediction. Broadband activity and alpha power produce different low-dimensional state-space geometries, while participant-specific state spaces can be mapped into a common coordinate system for between-participant comparison. In this exploratory analysis, aligned PCA decoded better early than either sensor space or shared PCA, whereas explicit trajectory descriptors performed poorly alone and added no reliable benefit when combined with PCA.

## 7 Beyond PCA

| Method | Primary Goal | Relation to PCA | Suitable for Trajectories? |
|---|---|---|---|
| PCA (Principal Component Analysis) | Identify orthogonal directions that capture maximal variance | ______________ | Yes - Well suited to time-resolved neural trajectories |
| ICA (Independent Component Analysis) | Separate signals into statistically independent components | Linear decomposition that optimizes statistical independence rather than variance; widely used in EEG/MEG for source separation and artifact identification | Yes - Trajectories can be constructed from time-resolved component activations, although ICA is not optimized specifically for trajectory geometry |
| FA (Factor Analysis) | Estimate principal components while modeling noise explicitly | Similar to PCA, but estimates principal components while explicitly modeling observation noise. | Yes - Especially useful for noisy trial-level data |
| dPCA (Demixed Principal Component Analysis) | Separate neural variance linked to specific task variables | PCA-like decomposition that uses task-variable labels to demix variance into interpretable components. | Excellent — ideal for task-related neural trajectories |
| LDA (Linear Discriminant Analysis) | Find dimensions that maximally discriminate classes | Alternative linear projection emphasizing separation rather than variance | Useful - Good for decoding and condition-specific subspaces |
| CCA (Canonical Correlation Analysis) | Identify maximally correlated linear projections between two datasets | Like PCA, provides linear projections of multivariate data, but optimizes correlation between paired datasets rather than variance within one dataset. | Yes - Useful for comparing or aligning trajectories across paired datasets or modalities |
| MDS (Multidimensional Scaling) | Represent pairwise distances or dissimilarities in a low-dimensional embedding | With Euclidean distances, classical MDS is equivalent to PCA; with other dissimilarities, MDS preserves pairwise structure rather than optimizing variance directly. | Yes - Useful when pairwise relationships among neural states are of primary interest, but temporal ordering is not explicitly modeled. |
| UMAP / t-SNE / PHATE / Isomap | Reveal nonlinear manifold structure | Nonlinear alternatives that preserve local, global, diffusion-based, or geodesic structure rather than maximizing linear variance. | Yes - But trajectory interpretation requires caution because temporal spacing and global geometry may be distorted. |

**Table 3 | PCA and its relatives: Overview of dimensionality-reduction, decomposition, and low-dimensional representation methods relevant to neural trajectory analysis.** These methods differ in their objectives: some preserve variance, others estimate components, separate task variables, maximize class separation, align datasets, or reveal nonlinear structure. The table is intended to help readers situate PCA within a broader landscape of available tools and choose dimensionality reduction methods according to their scientific question.

PCA is a useful entry point, but it is not the only method for obtaining and studying low-dimensional neural dynamics. As summarized in Table 3, different approaches become preferable when the goal is to separate task variables, preserve multiway structure, model low-dimensional dynamics, or identify nonlinear geometry.

Demixed PCA is useful when the experimental design contains known task variables. Standard PCA identifies axes that explain maximal variance, but these axes may mix stimulus identity, decision category, movement, time, and response preparation. Demixed PCA partitions variance according to task-related marginalizations before identifying low-dimensional axes within each aspect of the data (Kobak et al., 2016). For EEG and MEG, this can help determine whether trajectory geometry is shaped by stimulus encoding, decision formation, motor preparation, or time-dependent activity.

Although PCA provides a low-dimensional representation of the measured activity, it does not explicitly specify a generative model of the observable data or a noise model, distinguishing it from approaches such as factor analysis, Gaussian Process Factor Analysis (GPFA), and linear dynamical systems. Factor analysis and related generative methods are useful when trial-level data are noisy and the goal is to estimate shared low-dimensional structure while explicitly modeling noise. GPFA extends this logic by estimating smooth low-dimensional trajectories, particularly useful when the underlying neural state is expected to evolve smoothly over time (Yu et al., 2009). Linear dynamical systems go further by modeling how low-dimensional states evolve according to temporal rules, particularly useful when the scientific question concerns the dynamics themselves (Paninski et al., 2010; Linderman et al., 2017).

Tensor decompositions are useful when the multiway structure of EEG/MEG data should be preserved. Classical PCA requires flattening the data into a two-dimensional matrix, usually channels and time. This formulation can obscure structure across trials, conditions, frequencies, or subjects. Tensor methods factorize the data across several features simultaneously and can identify patterns shared across participants, specific to conditions recurrent across trials, or expressed in specific frequency bands (Acar et al., 2007; Williams et al., 2018).

Nonlinear methods such as Isomap, UMAP, PHATE, and t-SNE can reveal curved, branched, or clustered structure that linear PCA may miss (Tenenbaum et al., 2000; van der Maaten & Hinton, 2008; McInnes et al., 2018; Moon et al., 2019). However, they require caution for trajectory analysis. Many nonlinear embeddings do not preserve temporal order, global geometry, or continuous distances in a directly interpretable way. They should therefore be treated as exploratory tools and compared against simpler linear baselines when trajectory geometry is central to the interpretation.

The broader point is that PCA should be treated as a baseline and a bridge. It is simple enough to make state-space analysis accessible, but its linear formulation also exposes limits that motivate more structured methods. In particular, PCA does not explicitly model nonlinear or multiplicative interactions between components. Such interactions are common in neural representations; for example, gain-field responses in parietal cortex can combine spatial or movement-related signals with eye, hand, or body position through multiplicative modulation (Salinas & Abbott, 1996). In any case, PCA is always a good starting point, provided its limitations are taken into account when interpreting PCA results.

## 8 Opportunities, challenges, and practical recommendations

Low-dimensional trajectory analysis offers several opportunities for EEG and MEG research. It provides a framework for integrating ERPs, oscillatory activity, topographies, source estimates, connectivity patterns, and decoding results as different expressions of an evolving neural state. It may also help connect human electrophysiology with systems neuroscience by allowing state-space ideas developed in invasive recordings to be adapted to macroscopic signals. Finally, trajectory features such as speed, separation, stability, dimensionality, and subspace alignment may eventually contribute to studies of learning, aging, clinical variability, neurofeedback, or brain–computer interfaces.

An important challenge is avoiding overly literal interpretations of PCA geometry. Because PCA identifies orthogonal directions of decreasing variance, the structure of any given component depends on the variance already captured by preceding components. Higher-order PCs should therefore be interpreted in the context of the components that precede them rather than as independent neural processes. Moreover, potentially misleading dynamical structure can emerge from the projection itself. For example, smooth signals or similar response profiles shifted in time can produce principal components that result in apparent rotational trajectories even when the original signals contain no corresponding oscillatory dynamics (Shinn, 2023). Consequently, features such as rotations, loops, or other trajectory geometries should not by themselves be taken as evidence for an underlying dynamical mechanism, but should be evaluated against the structure of the original signals and appropriate control analyses.

These opportunities come with methodological challenges. The first is spatial mixing. EEG and MEG sensors do not measure isolated neural populations; they capture overlapping activity from distributed sources (Baillet, 2017; Gramfort et al., 2014). Sensor-space trajectories can summarize reproducible multichannel structure, but their coordinates reflect mixtures of activity from distributed neural sources rather than anatomically localized neural generators. Source reconstruction, spatial filtering, and ROI-based analyses can improve anatomical specificity, but introduce their own assumptions.

The second challenge is preprocessing sensitivity. Because PCA is variance-driven, residual artifacts can dominate the PC state space. Eye movements, muscle activity, slow drifts, line noise, filtering effects, baseline offsets, or high-amplitude sensors can shape the principal axes and, consequently, the resulting trajectories (Artoni et al., 2018; Shinn, 2023). Baseline correction is therefore an important preprocessing decision in PCA-based trajectory analysis. Researchers should specify whether baseline correction was applied, whether it was performed at the sensor, source, feature, or trial level, and which time window was used. Because baseline correction can alter the variance and covariance structure of the data, different baseline choices may affect the resulting PCA space and trajectories. When baseline activity may differ across conditions, researchers should therefore assess whether their results are robust to alternative baseline windows or to omitting baseline correction. More generally, researchers should inspect component loadings, report preprocessing choices clearly, and test whether results are robust across reasonable alternatives.

The third challenge is cross-subject comparison. PCA axes depend on the data used for fitting. If PCA is fit separately for each subject, alignment approaches such as Canonical Correlation Analysis (CCA) can help identify corresponding dimensions across subject- or session-specific representations and map them into a shared space (Gallego et al., 2020; Safaie et al., 2023; Zhang et al., 2017). Conversely, fitting PCA at the group level provides a common coordinate system but may obscure individual-specific structure. Studies should therefore report whether PCA was fitted separately or jointly across participants, how subject-specific representations were aligned, and how sign indeterminacy was handled. When analyses involve prediction or confirmatory evaluation, all data-dependent transformations, including PCA and alignment, should be estimated using the training data and subsequently applied to the held-out data to prevent information leakage and circularity (Varoquaux et al., 2017).

The fourth challenge is interpretation. Trajectory geometry is not self-explanatory. Separation, curvature, loops, convergence, or apparent attractor-like patterns may be meaningful, but they can also arise from projection choices, artifacts, oscillations, baseline structure, or filtering effects. Interpretation should therefore be anchored in task timing, behavior, component patterns, robustness analyses, and complementary methods.

A fifth challenge is assessing the quality of the dimensionality reduction itself. PCA trajectory studies should not only quantify the geometry of trajectories after projection but also evaluate whether the reduced dimensionality space provides an adequate representation of the original data structure. For PCA, this can include reporting explained variance, reconstruction error, residual variance, component stability across resampling, and the robustness of loadings or subspaces across subjects, sessions, or preprocessing choices. When the goal is to preserve relative distances or neighborhood structure, additional diagnostics such as Shepard diagrams, distance correlations between the full dimensionality and reduced dimensionality spaces, trustworthiness, and continuity can help assess whether the low-dimensional embedding distorts the relationships that are later interpreted as trajectory geometry. These diagnostics are especially important when PCA is compared with, or replaced by, nonlinear embedding methods.

A practical reporting checklist for PCA-based EEG/MEG trajectory analysis should therefore include: define the neural state; report preprocessing, scaling, and baseline correction; specify the data matrix; state which data was used for the PCA fit; report explained variance, reconstruction error, and component patterns; assess the quality and stability of the dimensionality reduction; distinguish visualization dimensions from inference dimensions; inspect artifacts; quantify trajectory geometry; validate against null models or held-out data; and relate trajectory features to task structure or behavior. These practices would make PCA trajectory analyses more reproducible, interpretable, reliable, and comparable across studies.

## 9 Conclusion

Low-dimensional trajectory analysis changes how EEG and MEG activity are conceptualized. Instead of treating electrophysiological recordings only as collections of channels, components, frequencies, or time windows, it represents each moment of activity as a multivariate state and asks how that state evolves over time. This shift in how one looks at data can reveal forms of organization that are difficult to capture with conventional analyses alone, including choice-dependent paths, attentional fluctuations, evolving sensory evidence, condition divergence, state convergence, and behaviorally relevant trajectory geometry.

PCA provides a practical and interpretable entry point into this framework; it offers a transparent coordinate system for summarizing dominant covariance structure and visualizing how distributed electrophysiological activity unfolds over time. When combined with component inspection, trajectory metrics, statistical validation, and behavioral anchoring, PCA can bridge classical human electrophysiology and population-dynamics approaches in systems neuroscience.

The future of PCA-based EEG/MEG state-space analysis will depend on careful state definition, transparent reporting, and strong validation. Low-dimensional representations become meaningful only when their components, scores, and trajectory geometry are linked to experimental design, behavior, component structure, robustness analyses, or computational models. Used in this way, PCA-based trajectory analysis does more than reduce dimensionality. It provides a shared geometric language for studying cognition as a dynamical process and for connecting human electrophysiology with broader principles of neural population dynamics.

## 10 Companion code, tutorials, and interactive reports

All code, tutorials, and analysis reports are openly available at

https://github.com/thecocolab/pca-neural-trajectories-eeg-meg.

***Tutorials.*** Six Jupyter notebooks in tutorials/ present the workflow as a graded, self-contained series intended to be didactic rather than merely reproducible. Each interleaves several explanatory sections with the code, motivating every step, stating the choices it commits to, and showing its output inline, so a reader can follow the reasoning without executing anything. Together they cover every analysis reported in this paper, panel by panel. On the EEG side, tutorial_eegbci_main.ipynb presents the example shown in Figure 2 for the four motor execution and imagery conditions, and tutorial_eegbci_decoding.ipynb covers the corresponding subject-disjoint decoding. On the MEG side, tutorial_megfaces_main.ipynb builds the shared broadband space of Figure 3a–b, tutorial_megfaces_spectral_envelopes.ipynb produces the alpha-band analysis of Figure 3c–d (and extends it to beta and low-gamma envelopes), and tutorial_megfaces_decoding.ipynb develops the cross-participant decoding and alignment workflow of Figure 3e–f. One notebook goes beyond the paper: tutorial_eegbci_nonlinear.ipynb compares PCA with UMAP, PHATE, and Isomap on the same observations. All notebooks include their outputs, so the figures and tables are visible on GitHub without running the code.

***Analysis scripts.*** Each notebook has a companion in scripts/ that applies the same workflow to the full cohort and records run provenance. Every script writes its tables, time-resolved metrics, fitted reducers, arrays, and figures, alongside a self-contained HTML report.

***Interactive reports*.** The generated reports are published at

https://thecocolab.github.io/pca-neural-trajectories-eeg-meg/.

A single index links every analysis for both datasets. Each report is a standalone HTML page with interactive figures, requiring no installation.

***Data*.** The MEG data are the Wakeman–Henson multimodal dataset, OpenNeuro ds000117 (Wakeman & Henson, 2015). The EEG data are the PhysioNet EEG Motor Movement/Imagery Database. Neither is redistributed here; the notebooks fetch and preprocess them explicitly.

## Acknowledgments

V.H. is supported by funding from the Canadian Institutes of Health Research (CIHR; PJT-166197). P.C. is supported by the Natural Sciences and Engineering Research Council of Canada (NSERC; RGPIN-2022-05345) and the Canadian Institutes of Health Research (CIHR; PJT-198190). K.J. is supported by funding from the Canada Research Chairs (950-232368) program and a Discovery Grant from the Natural Sciences and Engineering Research Council of Canada (2021-03426).

## References

Acar, E., Aykut-Bingöl, C., Bingöl, H., Bro, R., & Yener, B. (2007). Multiway analysis of epilepsy tensors. *Bioinformatics, 23*(13), i10–i18. https://doi.org/10.1093/bioinformatics/btm210

Arcara, G., Pellegrino, G., Pascarella, A., Mantini, D., Kobayashi, E., & Jerbi, K. (2023). MEG. In *Psychophysiology methods* (pp. 157–180). Springer.

Artoni, F., Delorme, A., & Makeig, S. (2018). Applying dimension reduction to EEG data by principal component analysis reduces the quality of its subsequent independent component decomposition. *NeuroImage, 175*, 176–187. https://doi.org/10.1016/j.neuroimage.2018.03.016

Baillet, S. (2017). Magnetoencephalography for brain electrophysiology and imaging. *Nature Neuroscience, 20*, 327–339. https://doi.org/10.1038/nn.4504

Brkić, D., Sommariva, S., Schuler, A. L., Pascarella, A., Belardinelli, P., Isabella, S. L., & Pellegrino, G. (2023). The impact of ROI extraction method for MEG connectivity estimation: Practical recommendations for the study of resting state data. *NeuroImage, 284*, 120424. https://doi.org/10.1016/j.neuroimage.2023.120424

Chung, S., & Abbott, L. F. (2021). Neural population geometry: An approach for understanding biological and artificial neural networks. *Current Opinion in Neurobiology, 70*, 137–144. https://doi.org/10.1016/j.conb.2021.10.010

Churchland, M. M., Cunningham, J. P., Kaufman, M. T., Foster, J. D., Nuyujukian, P., Ryu, S. I., & Shenoy, K. V. (2012). Neural population dynamics during reaching. *Nature, 487*, 51–56. https://doi.org/10.1038/nature11129

Cleasby, I. R., Wakefield, E. D., Morrissey, B. J., Bodey, T. W., Votier, S. C., Bearhop, S., & Hamer, K. C. (2019). Using time-series similarity measures to compare animal movement trajectories in ecology. *Behavioral Ecology and Sociobiology, 73*, 151. https://doi.org/10.1007/s00265-019-2761-1

Cohen, M. X. (2014). *Analyzing neural time series data: Theory and practice*. MIT Press.

Colclough, G. L., Woolrich, M. W., Tewarie, P. K., Brookes, M. J., Quinn, A. J., & Smith, S. M. (2016). How reliable are MEG resting-state connectivity metrics? *NeuroImage, 138*, 284–293. https://doi.org/10.1016/j.neuroimage.2016.05.070

Cunningham, J. P., & Yu, B. M. (2014). Dimensionality reduction for large-scale neural recordings. *Nature Neuroscience, 17*, 1500–1509. https://doi.org/10.1038/nn.3776

de Cheveigné, A., & Simon, J. Z. (2007). Denoising based on time-shift PCA. *Journal of Neuroscience Methods, 165*(2), 297–305. https://doi.org/10.1016/j.jneumeth.2007.06.003

Dekker, M. M., França, A. S., Panja, D., & Cohen, M. X. (2021). Characterizing neural phase-space trajectories via principal Louvain clustering. *Journal of Neuroscience Methods, 362*, 109313. https://doi.org/10.1016/j.jneumeth.2021.109313

Demšar, U., Buchin, K., Cagnacci, F., Safi, K., Speckmann, B., Van de Weghe, N., Weiskopf, D., & Weibel, R. (2015). Analysis and visualisation of movement: An interdisciplinary review. *Movement Ecology, 3*, 5. https://doi.org/10.1186/s40462-015-0032-y

Do, J., James, O., & Kim, Y. J. (2024). Choice-dependent delta-band neural trajectory during semantic category decision making in the human brain. *iScience, 27*(7), 110173. https://doi.org/10.1016/j.isci.2024.110173

Durbin, J., & Koopman, S. J. (2012). *Time series analysis by state space methods* (2nd ed.). Oxford University Press.

Elsayed, G. F., Lara, A. H., Kaufman, M. T., Churchland, M. M., & Cunningham, J. P. (2016). Reorganization between preparatory and movement population responses in motor cortex. *Nature Communications, 7,* 13239. https://doi.org/10.1038/ncomms13239

Gallego, J. A., Perich, M. G., Chowdhury, R. H., Solla, S. A., & Miller, L. E. (2020). Long-term stability of cortical population dynamics underlying consistent behavior. *Nature Neuroscience, 23*, 260–270. https://doi.org/10.1038/s41593-019-0555-4

Gallego, J. A., Perich, M. G., Miller, L. E., & Solla, S. A. (2017). Neural manifolds for the control of movement. *Neuron, 94*(5), 978–984. https://doi.org/10.1016/j.neuron.2017.05.025

Gallego, J. A., Perich, M. G., Naufel, S. N., Ethier, C., Solla, S. A., & Miller, L. E. (2018). Cortical population activity within a preserved neural manifold underlies multiple motor behaviors. *Nature Communications, 9*, 4233. https://doi.org/10.1038/s41467-018-06560-z

Goldberger, A. L., Amaral, L. A. N., Glass, L., Hausdorff, J. M., Ivanov, P. Ch., Mark, R. G., Mietus, J. E., Moody, G. B., Peng, C.-K., & Stanley, H. E. (2000). PhysioBank, PhysioToolkit, and PhysioNet: Components of a new research resource for complex physiologic signals. *Circulation, 101*(23), e215–e220. https://doi.org/10.1161/01.CIR.101.23.e215

Gramfort, A., Luessi, M., Larson, E., Engemann, D. A., Strohmeier, D., Brodbeck, C., Parkkonen, L., & Hämäläinen, M. S. (2013). MEG and EEG data analysis with MNE-Python. *Frontiers in Neuroscience, 7*, 267. https://doi.org/10.3389/fnins.2013.00267

Gramfort, A., Luessi, M., Larson, E., Engemann, D. A., Strohmeier, D., Brodbeck, C., Goj, R., Jas, M., Brooks, T., Parkkonen, L., & Hämäläinen, M. S. (2014). MNE software for processing MEG and EEG data. *NeuroImage, 86*, 446–460. https://doi.org/10.1016/j.neuroimage.2013.10.027

Greenacre, M., Groenen, P.J.F., Hastie, T. et al. Principal component analysis. Nat Rev Methods Primers 2, 100 (2022). https://doi.org/10.1038/s43586-022-00184-w

Heller, C. R., & David, S. V. (2022). Targeted dimensionality reduction enables reliable estimation of neural population coding accuracy from trial-limited data. *PLOS ONE, 17*(7), e0271136. https://doi.org/10.1371/journal.pone.0271136

Hyvärinen, A., & Oja, E. (2000). Independent component analysis: Algorithms and applications. *Neural Networks, 13*(4–5), 411–430. https://doi.org/10.1016/S0893-6080(00)00026-5

Jazayeri, M., & Ostojic, S. (2021). Interpreting neural computations by examining intrinsic and embedding dimensionality of neural activity. *Current Opinion in Neurobiology, 70*, 113–120. https://doi.org/10.1016/j.conb.2021.08.002

Jolliffe, I. T., & Cadima, J. (2016). Principal component analysis: A review and recent developments. *Philosophical Transactions of the Royal Society A, 374*(2065), 20150202. https://doi.org/10.1098/rsta.2015.0202

Jung, T.-P., Makeig, S., McKeown, M. J., Bell, A. J., Lee, T.-W., & Sejnowski, T. J. (2001). Imaging brain dynamics using independent component analysis. *Proceedings of the IEEE, 89*(7), 1107–1122. https://doi.org/10.1109/5.939827

Kaufman, M. T., Churchland, M. M., Ryu, S. I., & Shenoy, K. V. (2014). Cortical activity in the null space: Permitting preparation without movement. *Nature Neuroscience, 17*, 440–448. https://doi.org/10.1038/nn.3643

Kayser, J., & Tenke, C. E. (2003). Optimizing PCA methodology for ERP component identification and measurement: Theoretical rationale and empirical evaluation. *Clinical Neurophysiology, 114*(12), 2307–2325. https://doi.org/10.1016/S1388-2457(03)00241-4

Keil, A., Debener, S., Gratton, G., Junghöfer, M., Kappenman, E. S., Luck, S. J., Luu, P., Miller, G. A., & Yee, C. M. (2014). Committee report: Publication guidelines and recommendations for studies using electroencephalography and magnetoencephalography. *Psychophysiology, 51*(1), 1–21. https://doi.org/10.1111/psyp.12147

Kobak, D., Brendel, W., Constantinidis, C., Feierstein, C. E., Kepecs, A., Mainen, Z. F., Romo, R., Qi, X.-L., Uchida, N., & Machens, C. K. (2016). Demixed principal component analysis of neural population data. *eLife, 5*, e10989. https://doi.org/10.7554/eLife.10989

Langdon, C., Genkin, M., & Engel, T. A. (2023). A unifying perspective on neural manifolds and circuits for cognition. *Nature Reviews Neuroscience, 24*, 363–377. https://doi.org/10.1038/s41583-023-00693-x

Lee, F., Scherer, R., Leeb, R., Schlögl, A., Bischof, H., & Pfurtscheller, G. (2004). Feature mapping using PCA, locally linear embedding and isometric feature mapping for EEG-based brain computer interface. In *Proceedings of the 2004 IEEE International Joint Conference on Neural Networks* (pp. 189–196).

Linderman, S. W., Johnson, M. J., Miller, A. C., Adams, R. P., Blei, D. M., & Paninski, L. (2017). Bayesian learning and inference in recurrent switching linear dynamical systems. Proceedings of the 20th International Conference on Artificial Intelligence and Statistics, Proceedings of Machine Learning Research, 54, 914–922.

Liu, J., Harris, A., & Kanwisher, N. (2002). Stages of processing in face perception: An MEG study. *Nature Neuroscience, 5*(9), 910–916. https://doi.org/10.1038/nn909

Machens, C. K., Romo, R., & Brody, C. D. (2010). Functional, but not anatomical, separation of “what” and “when” in prefrontal cortex. *Journal of Neuroscience, 30*(1), 350–360. https://doi.org/10.1523/JNEUROSCI.3276-09.2010

Maheswaranathan, N., Williams, A. H., Golub, M. D., Ganguli, S., & Sussillo, D. (2019). Reverse engineering recurrent networks for sentiment classification reveals line attractor dynamics. *Advances in Neural Information Processing Systems, 32*.

Makeig, S., Bell, A. J., Jung, T.-P., & Sejnowski, T. J. (1996). Independent component analysis of electroencephalographic data. *Advances in Neural Information Processing Systems, 8*, 145–151.

Mante, V., Sussillo, D., Shenoy, K. V., & Newsome, W. T. (2013). Context-dependent computation by recurrent dynamics in prefrontal cortex. *Nature, 503*, 78–84. https://doi.org/10.1038/nature12742

McInnes, L., Healy, J., & Melville, J. (2018). UMAP: Uniform manifold approximation and projection for dimension reduction. *arXiv*. https://doi.org/10.48550/arXiv.1802.03426

Mitchell-Heggs, R., Prado, S., Gava, G. P., Go, M. A., & Schultz, S. R. (2023). Neural manifold analysis of brain circuit dynamics in health and disease. *Journal of Computational Neuroscience, 51*, 1–21. https://doi.org/10.1007/s10827-022-00839-3

Moon, K. R., van Dijk, D., Wang, Z., Gigante, S., Burkhardt, D. B., Chen, W. S., Yim, K., Elzen, A. van den, Hirn, M. J., Coifman, R. R., Ivanova, N. B., Wolf, G., & Krishnaswamy, S. (2019). Visualizing structure and transitions in high-dimensional biological data. *Nature Biotechnology, 37*, 1482–1492. https://doi.org/10.1038/s41587-019-0336-3

Morales, S., & Bowers, M. E. (2022). Time-frequency analysis methods and their application in developmental EEG data. *Developmental Cognitive Neuroscience, 54*, 101067. https://doi.org/10.1016/j.dcn.2022.101067

Oby, E. R., Degenhart, A. D., Grigsby, E. M., Motiwala, A., McClain, N. T., Marino, P. J., Yu, B. M., & Batista, A. P. (2025). Dynamical constraints on neural population activity. *Nature Neuroscience, 28*, 383–393. https://doi.org/10.1038/s41593-024-01845-7

Paninski, L., Ahmadian, Y., Ferreira, D. G., Koyama, S., Rad, K. R., Vidne, M., Vogelstein, J., & Wu, W. (2010). A new look at state-space models for neural data. *Journal of Computational Neuroscience, 29*, 107–126. https://doi.org/10.1007/s10827-009-0179-x

Pellegrini, F., Delorme, A., Nikulin, V., & Haufe, S. (2023). Identifying good practices for detecting inter-regional linear functional connectivity from EEG. *NeuroImage, 277*, 120218. https://doi.org/10.1016/j.neuroimage.2023.120218

Recanatesi, S., Bradde, S., Balasubramanian, V., Steinmetz, N. A., & Shea-Brown, E. (2022). A scale-dependent measure of system dimensionality. *Patterns, 3*(8), 100555. https://doi.org/10.1016/j.patter.2022.100555

Rossion, B., Joyce, C. A., Cottrell, G. W., & Tarr, M. J. (2003). Early lateralization and orientation tuning for face, word, and object processing in the visual cortex. NeuroImage, 20(3), 1609–1624. https://doi.org/10.1016/j.neuroimage.2003.07.010

Russo, A. A., Khajeh, R., Bittner, S. R., Perkins, S. M., Cunningham, J. P., Abbott, L. F., & Churchland, M. M. (2020). Neural trajectories in the supplementary motor area and motor cortex exhibit distinct geometries, compatible with different classes of computation. *Neuron, 107*(4), 745–758.e6. https://doi.org/10.1016/j.neuron.2020.05.020

Saeidi, M., Karwowski, W., Farahani, F. V., Fiok, K., Taiar, R., Hancock, P. A., & Al-Juaid, A. (2021). Neural decoding of EEG signals with machine learning: A systematic review. *Brain Sciences, 11*(11), 1525. https://doi.org/10.3390/brainsci11111525

Safaie, M., Chang, J. C., Park, J., Miller, L. E., Dudman, J. T., Perich, M. G., & Gallego, J. A. (2023). Preserved neural dynamics across animals performing similar behaviour. *Nature, 623*(7988), 765–771. https://doi.org/10.1038/s41586-023-06714-0

Salinas, E., & Abbott, L. F. (1996). A model of multiplicative neural responses in parietal cortex. *Proceedings of the National Academy of Sciences, 93*(21), 11956–11961. https://doi.org/10.1073/pnas.93.21.11956

Schalk, G., McFarland, D. J., Hinterberger, T., Birbaumer, N., & Wolpaw, J. R. (2004). BCI2000: A general-purpose brain-computer interface (BCI) system. *IEEE Transactions on Biomedical Engineering, 51*(6), 1034–1043. https://doi.org/10.1109/TBME.2004.827072

Seidel, D. P., Dougherty, E., Carlson, C., & Getz, W. M. (2018). Ecological metrics and methods for GPS movement data. *International Journal of Geographical Information Science, 32*(11), 2272–2293. https://doi.org/10.1080/13658816.2018.1498097

Shenoy, K. V., Sahani, M., & Churchland, M. M. (2013). Cortical control of arm movements: A dynamical systems perspective. *Annual Review of Neuroscience, 36*, 337–359. https://doi.org/10.1146/annurev-neuro-062111-150509

Shinn, M. (2023). Phantom oscillations in principal component analysis. *Proceedings of the National Academy of Sciences, 120*(48), e2311420120. https://doi.org/10.1073/pnas.2311420120

Song, H., Shim, W. M., & Rosenberg, M. D. (2023). Large-scale neural dynamics in a shared low-dimensional state space reflect cognitive and attentional dynamics. *eLife, 12*, e85487. https://doi.org/10.7554/eLife.85487

Sussillo, D. (2014). Neural circuits as computational dynamical systems. *Current Opinion in Neurobiology, 25*, 156–163. https://doi.org/10.1016/j.conb.2014.01.013

Sussillo, D., & Barak, O. (2013). Opening the black box: Low-dimensional dynamics in high-dimensional recurrent neural networks. *Neural Computation, 25*(3), 626–649. https://doi.org/10.1162/NECO_a_00409

Tao, Y., Both, A., Silveira, R. I., Buchin, K., Sijben, S., Purves, R. S., Laube, P., Peng, D., Toohey, K., & Duckham, M. (2021). A comparative analysis of trajectory similarity measures. *GIScience & Remote Sensing, 58*(5), 643–669. https://doi.org/10.1080/15481603.2021.1908927

Tenenbaum, J. B., de Silva, V., & Langford, J. C. (2000). A global geometric framework for nonlinear dimensionality reduction. *Science, 290*(5500), 2319–2323. https://doi.org/10.1126/science.290.5500.2319

Thiery, T., Rainville, P., Cisek, P., & Jerbi, K. (2022). Distinct trajectories in low-dimensional neural oscillation state space track dynamic decision-making in humans. *bioRxiv*. https://doi.org/10.1101/2022.06.14.494674

Thura, D., Cabana, J. F., Feghaly, A., & Cisek, P. (2022). Integrated neural dynamics of sensorimotor decisions and actions. *PLOS Biology, 20*(12), e3001861. https://doi.org/10.1371/journal.pbio.3001861

Tyagi, A., & Nehra, V. (2017). A comparison of feature extraction and dimensionality reduction techniques for EEG-based BCI system. *IUP Journal of Computer Sciences, 11*(1).

Urigüen, J. A., & Garcia-Zapirain, B. (2015). EEG artifact removal: State-of-the-art and guidelines. *Journal of Neural Engineering, 12*(3), 031001. https://doi.org/10.1088/1741-2560/12/3/031001

Uusitalo, M. A., & Ilmoniemi, R. J. (1997). Signal-space projection method for separating MEG or EEG into components. *Medical and Biological Engineering and Computing, 35*(2), 135–140. https://doi.org/10.1007/BF02534144

van der Maaten, L., & Hinton, G. (2008). Visualizing data using t-SNE. *Journal of Machine Learning Research, 9*, 2579–2605.

Varoquaux, G., Raamana, P. R., Engemann, D. A., Hoyos-Idrobo, A., Schwartz, Y., & Thirion, B. (2017). Assessing and tuning brain decoders: Cross-validation, caveats, and guidelines. *NeuroImage, 145*, 166–179. https://doi.org/10.1016/j.neuroimage.2016.10.038

Vidaurre, D., Hunt, L. T., Quinn, A. J., Hunt, B. A. E., Brookes, M. J., Nobre, A. C., & Woolrich, M. W. (2018). Spontaneous cortical activity transiently organises into frequency-specific phase-coupling networks. *Nature Communications, 9*, 2987. https://doi.org/10.1038/s41467-018-05316-z

von Wegner, F., Knaut, P., & Laufs, H. (2018). EEG microstate sequences from different clustering algorithms are information-theoretically invariant. *Frontiers in Computational Neuroscience, 12*, 70. https://doi.org/10.3389/fncom.2018.00070

Vyas, S., Golub, M. D., Sussillo, D., & Shenoy, K. V. (2020). Computation through neural population dynamics. *Annual Review of Neuroscience, 43*, 249–275. https://doi.org/10.1146/annurev-neuro-092619-094115

Wakeman, D. G., & Henson, R. N. (2015). A multi-subject, multi-modal human neuroimaging dataset. Scientific Data, 2, 150001. https://doi.org/10.1038/sdata.2015.1

Williamson, R. C., Doiron, B., Smith, M. A., & Yu, B. M. (2019). Bridging large-scale neuronal recordings and large-scale network models using dimensionality reduction. *Current Opinion in Neurobiology, 55*, 40–47. https://doi.org/10.1016/j.conb.2018.12.008

Williams, A. H., Kim, T. H., Wang, F., Vyas, S., Ryu, S. I., Shenoy, K. V., Schnitzer, M., Kolda, T. G., & Ganguli, S. (2018). Unsupervised discovery of demixed, low-dimensional neural dynamics across multiple timescales through tensor component analysis. *Neuron, 98*(6), 1099–1115.e8. https://doi.org/10.1016/j.neuron.2018.05.015.

Yan, Y., Goodman, J. M., Moore, D. D., Solla, S. A., & Bensmaia, S. J. (2020). Unexpected complexity of everyday manual behaviors. *Nature Communications, 11*, 3564. https://doi.org/10.1038/s41467-020-17404-0

Yu, B. M., Cunningham, J. P., Santhanam, G., Ryu, S. I., Shenoy, K. V., & Sahani, M. (2009). Gaussian-process factor analysis for low-dimensional single-trial analysis of neural population activity. *Journal of Neurophysiology, 102*(1), 614–635. https://doi.org/10.1152/jn.90941.2008

Zhang, Q., Borst, J. P., Kass, R. E., & Anderson, J. R. (2017). Inter-subject alignment of MEG datasets in a common representational space. *Human Brain Mapping, 38*(9), 4287–4301. doi:10.1002/hbm.23689

Zhang, G., Carrasco, C. D., Winsler, K., Bahle, B., Cong, F., & Luck, S. J. (2024). Assessing the effectiveness of spatial PCA on SVM-based decoding of EEG data. *NeuroImage, 293*, 120625. https://doi.org/10.1016/j.neuroimage.2024.120625